\documentclass[aps,showpacs,floatfix,twocolumn]{revtex4}
\usepackage{graphicx}
\usepackage{amsmath}
\begin{document}
\title{Microscopic Study of (p,$\gamma$) Reactions in Mass Region A=110-125}
\author{Dipti Chakraborty}
\email{diptichakraborty2011@gmail.com}
\author{Saumi Dutta}
\email{saumidutta89@gmail.com}
\author{G. Gangopadhyay}
\email{ggphy@caluniv.ac.in}
\author{Abhijit Bhattacharyya}
\email{abhattacharyyacu@gmail.com}
\affiliation{Department of Physics, University of Calcutta, \\92 Acharya Prafulla Chandra Road, Kolkata-700009}

\begin{abstract}
Low energy proton capture reactions have been studied in statistical model
using the semi-microscopic optical potential
in the mass range 
A=110-125. Nuclear density obtained from relativistic mean field calculation 
has been folded with nucleon-nucleon interaction to obtain the optical 
potential. Theoretical results have been compared with experimental
measurements to normalize the potential. Results have also been compared 
with a standard calculation available in the literature.
\end{abstract}
\pacs{24.10.Ht,25.40.Lw,25.40.Kv}
\maketitle

Creation of nuclei heavier than iron occurs mainly through three processes - 
slow and rapid neutron capture processes and p-process~\cite{burbidge,sprocess,rprocess,arnould}. 
Though some studies have shown that p-process 
amounts to only 0.1\% of the total abundance of nuclei heavier than iron, 
there are possibly 35 isotopes on the neutron deficient side of valley of 
stability expected to be produced either entirely or partly through the 
 p-process. The p-process depends on reactions such as ($p,\gamma$),
($\gamma,p$) 
 and ($\gamma,n$) reactions,  as well as on the seed nuclei.  

In spite of the high Coulomb barrier, various astrophysical sites may produce 
medium mass proton rich nuclei 
through radiative proton capture reaction with a huge proton flux.
One of them is the x-ray burst where 
the peak temperature reaches 1-3 GK. 
In heavier nuclei, it is important to study the ($\gamma,p$) reactions which 
destroy proton rich nuclei and compete with the ($\gamma,n$) reactions.
In terrestrial laboratories, it is easier to study $(p,\gamma)$ reactions
and infer the cross-sections for the inverse ($\gamma,p$) reactions.
Besides, in astrophysical environments where photodisintegration occurs,
the temperature is very high and
nuclei are likely to exist in excited states as distinct from our laboratory
experiments. Hence, the contribution of the ground state may not be substantial.
All these arguments have led to the study of the forward $(p,\gamma)$ reactions
to constrain the rates.

To study the actual abundance of different nuclei and evolution of the 
process, a network calculation involving a large 
number of nuclear reactions is required. The network calculation in an
explosive astrophysical situation has to take many quantities such as
temperature, pressure, proton mass fraction,  
forward and inverse reaction rates etc in to account.
Thus, we need to tune  the interaction potential properly.
As the p-process proceeds along proton rich side of the stability valley, it 
involves many nuclei which are unstable and inaccessible as targets on earth. 
Hence, theory remains the principal guide to gather information about various
reactions. It is useful to construct the nucleon-nucleus potential
through folding of the NN interaction with the
nuclear density.  Nuclear density 
for stable targets may be experimentally measured by electron scattering.
In unstable nuclei experimental nuclear densities are not available and 
theoretical results may be used.

Rauscher {\em et al.}~\cite{raus1,raus2} have calculated the reaction rates for various
proton, neutron and alpha induced reactions and their inverse reactions in 
Hauser-Feshbach formalism for various targets over wide ranges of atomic 
numbers, masses and temperatures in global approach. 
Optical potential, a key ingredient for Hauser-Feshbach calculation, is often 
taken in a local or global approach.
In the present work we concentrate on the nuclei in the mass region 
A=110-125.
There are a number of p-nuclei in and around this region, {\em viz.} 
$^{113}$In, $^{112,114,115}$Sn, $^{120}$Te and $^{124,126}$Xe.
Away from the stability valley, reaction rates calculated even from 
semi-microscopic optical potential and phenomenological density prescription,
become very uncertain and have to be varied by a large amount~\cite{schatz}. 
We have used a more microscopic approach  where an appropriate microscopic NN 
interaction has been folded with 
 nuclear density from mean field calculations  to reduce this uncertainty.
This process can be extended to the unstable nuclei. 
It is imperative to test the theoretical method in reactions where
experimental data are available to verify its applicability before 
extending it to unknown isotopes. 
Hence, reaction cross-section for available stable targets in the mass region have 
been studied in the present work.

The density dependent M3Y interaction (DDM3Y), which is known to give satisfactory 
results in many cases, has been chosen for our purpose. Similar studies in mass 
region A=55-100~\cite{gg,chi1,chi2,chi3,saumi} have already been carried out 
and our aim is to extend this approach to A=110-125 region. 

Energy being low in astrophysical environment, the nuclear skin plays a 
very important role in the nuclear reactions. In astrophysical situation 
nuclear reaction rate is sensitively dependent on 
nuclear density profile. Reliable density information can be obtained from 
relativistic mean 
field (RMF) theory, a very useful tool for explaining various low energy nuclear properties and nuclear 
structure as well as nuclear density~\cite{muller,ring}.
    
Nuclear density profiles have been calculated using the Lagrangian density 
FSU Gold~\cite{routel}. This Lagrangian density contains, apart from the usual 
terms for 
nucleons and mesons, two additional non-linear 
meson-meson interaction terms, whose main role is to soften the EOS of symmetric
matter and reduce the symmetry energy. 

 Calculation of density has been carried out in 
co-ordinate space assuming spherical symmetry. A pairing force of 
300 MeV-fm for both protons and neutrons has been used under BCS 
approximation.  More details for the method are available in Ref.~\cite{bhattacharya}.  Considering the finite size of the proton, charge density is obtained by convoluting the point proton density.
\begin{equation}
\rho_{ch}({\mathbf r})=e \int \rho( {\mathbf r \prime})g({\mathbf r}-{\mathbf r\prime})d{\mathbf r\prime}
\end {equation}
Here, $g(\vec{r})$ is the Gaussian form factor given by,
\begin{equation}
  g(\vec{r})=(a\sqrt{\pi})^{-3} exp(-r^{2}/a^{2})
\end{equation}
where $a$ is a constant whose value is assigned to be 0.8 fm.

It is well known that RMF can provide an excellent description of nuclear 
ground state properties in this mass region.To check the validity of nuclear 
density calculation, nuclear charge radii may be compared with measured 
values. In table \ref{tab:exp} calculated binding 
energies and charge radii values have been compared with experimental 
measurements. The theoretical 
binding energy values have been corrected following the phenomenological 
procedure in Refs.~\cite{becor,becor1}. Here we have listed all the stable 
isotopes which occur in 
the concerned mass region and whose experimental charge radii values are known.

The relative difference between theoretical charge radii and experimental 
values is less than 0.5\%. 
 Binding energy values show a maximum error of 0.2\%.
  Experimental binding energy and charge radii have been taken from Audi {\em et al.}~\cite{audi} 
and  Angeli~\cite{radii}, respectively. Effect of  centre of mass correction 
would be small for heavy nuclei as relative correction in radius goes as 
A$^{-4/3}$, where A is the mass number of the nucleus~\cite{quentin}.

The M3Y interaction~\cite{bertsch,satchler}, even in absence of
explicit density and energy dependence, is able to produce reasonably satisfactory results for 
processes like elastic scattering and reaction.
After incorporation of density dependence, it is termed as density dependent 
M3Y (DDM3Y) interaction~\cite{myers}. The effective nucleon-nucleon interaction 
incorporates the density term as 
 \begin{equation}
g(\rho,\epsilon)=C(1-\beta(\epsilon)\rho^{2/3})
\end{equation}
$\rho$ and $\epsilon$ are the nucleonic density and energy per nucleon,
respectively.
The constants $C$ and $\beta$ have been assigned to be 2.07 and 
1.624 fm$^{2}$, respectively~\cite{dn1} obtained from optimum fit for $\alpha$ 
radioactivity data~\cite{dn2}. Here we also have used Scheerbaum 
spin orbit potential term~\cite{scheer} with phenomenological complex potential 
depths. The values used here for these phenomenological potential depths are 
same as in Refs.~\cite{gg,chi1,chi2,chi3}. 

The DDM3Y interaction has been incorporated in TALYS1.4 code~\cite{koning} for
Hauser-Feshbach statistical model calculations. We have used the Goriely 
HFB-Skyrme calculations 
for theoretical nuclear masses in TALYS code for the nuclei whose experimental 
masses are not available. The 
prescriptions for level densities and gamma 
strength function are taken as Constant Temperature Fermi Gas model and 
Goriely's Hybrid model~\cite{SGoriley}, respectively. All these features are
available in the code. We have included the effect of the width fluctuation 
correction which has a significant impact at low incident energies.
We have considered a maximum of 30 discrete levels in target nuclei and in 
residual nuclides. Full $j-l$ coupling has been used.

We have compared theoretical and experimental astrophysical S-factors. 
S-factor is a more slowly varying function of energy than the cross-section,  
and useful for comparison of results at very low energy.

The Gamow window for proton capture reaction is centered around the Gamow 
energy~\cite{rolfs} in MeV 
\begin{equation}
E_{0}=0.1220(Z_{t}^{2}Z_{p}^{2}\mu T_{9}^{2})^{1/3} 
\end{equation}
with width in MeV
\begin{equation}
\Delta = 0.2368 (Z_{p}^{2}Z_{t}^{2}\mu T_{9}^{5})^{1/6}
\end{equation}
Here, $Z_{p}$ and $Z_{t}$ are atomic numbers of the projectile and the target,  
$\mu$ is the reduced mass in a.m.u. of the system and T$_{9}$ is the temperature in GK.
In the reactions considered in the present work Gamow window has a range from 1.8 to  
4.5 MeV for temperature around 2-3 GK.
Experimental masses have been taken from Ref.~\cite{audi} in converting obtained reaction cross-section values to S-factor for comparison with experimental S-factor values.
For $(\gamma,p)$ reaction, it is worth noting that effective energy window will be centered at
\begin{equation} E^{eff}_\gamma=E_0+Q_{p\gamma}\end{equation}
because photodisintegration requires energy in excess of the threshold value,
keeping the width unchanged.

In astrophysical environments temperature may be high and nuclei can exist in 
ground state as well as various excited states. Hence, Maxwellian averaging has 
been performed for reaction rates.

\begin{table*}[tbh]
\center
\caption{\footnotesize Calculated binding energy (MeV) and charge radii (fm) of nuclei in A=110-125 
region compared with measured values.
\label{tab:exp}}
\begin{tabular}{ c rrrr| crrrr | crrrr} 
\hline
Nucleus     & \multicolumn{2}{c} {Binding energy}&   \multicolumn{2}{c|} {Charge radius} &
Nucleus     & \multicolumn{2}{c} {Binding energy}&   \multicolumn{2}{c|} {Charge radius} &
 Nucleus     & \multicolumn{2}{c} {Binding energy}&   \multicolumn{2}{c} {Charge radius} 
\\
  &   Theory   &      Expt.  &        Theory   &      Expt.  &
  &   Theory   &      Expt.  &        Theory   &      Expt.  &     
  & Theory   &      Expt.  &        Theory   &      Expt.  \\
\hline
$^{110}$Pd&   939.78     &  940.21   &4.552	&		4.578 &
$^{112}$Sn& 952.46  & 953.53  &     4.609 &         4.594 &
$^{124}$Sn& 1047.73 &  1049.96  &   4.695  &         4.676  \\
$^{110}$Cd& 939.24  &  940.64     &     4.575 &          4.574&
$^{114}$Sn& 970.64  & 971.57  &     4.622  &         4.610&
$^{121}$Sb& 1026.62 & 1026.32  &     4.690 &         4.680 \\
$^{111}$Cd& 947.98       &   947.62        &       4.581 &         4.580 &
$^{115}$Sn&  979.14 & 979.12  & 4.630      &         4.617 &
$^{123}$Sb& 1042.30 & 1042.10  &     4.704  &         4.688 \\
$^{112}$Cd& 956.38  &  957.01   &   4.588  &         4.595&
$^{116}$Sn& 987.24  & 988.68  &     4.638  &         4.627&
$^{122}$Te&1033.75&1034.33&4.712&4.708\\
$^{113}$Cd& 964.06 & 963.55   & 4.560         &    4.601  & 
$^{117}$Sn& 995.30 &    995.62 &        4.645&		4.632 &
$^{123}$Te& 1041.99& 1041.26   &	4.719&		4.711\\
$^{114}$Cd& 971.82 &    972.60&        4.604  &        4.614 &
$^{118}$Sn& 1003.27       &    1004.95  &	4.653 &		4.641&
$^{124}$Te&1050.13&1050.69&4.725&4.718\\
$^{116}$Cd& 986.47 &   987.43 &     4.619    &     4.628&
$^{119}$Sn& 1010.96 & 1011.43  &     4.660 &         4.645 &
$^{125}$Te&1058.16&1057.26&4.732&4.720 \\
$^{113}$In&  963.88 &  963.09    &     4.605 &          4.602 &
$^{120}$Sn& 1018.57& 1020.54   & 	4.667&		4.654&
$^{124}$Xe&1046.43         & 1046.26         &     4.755 &           4.762\\
$^{115}$In&   980.06&      979.40& 4.621 &       4.617 &
$^{122}$Sn& 1033.52& 1035.52   &	4.681 &		4.666 \\
\hline
\end{tabular}
\end{table*}

\begin {figure*}[htb]
\centering
\includegraphics[scale=0.25]{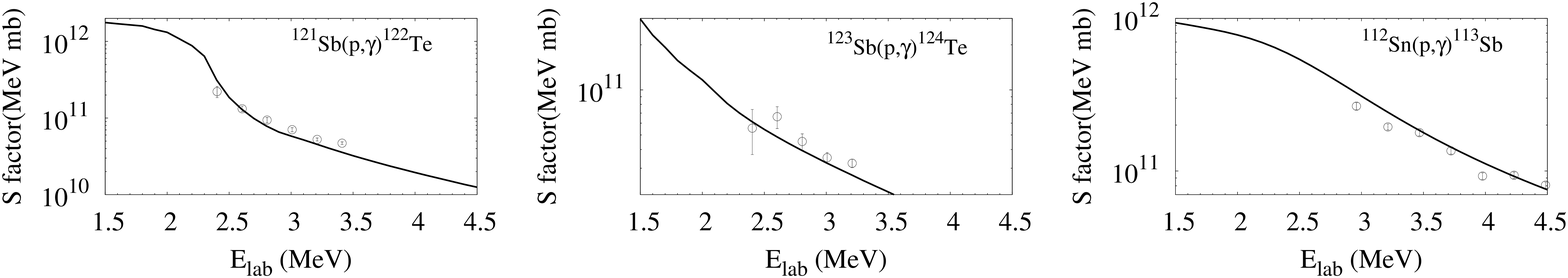}
\includegraphics[scale=0.25]{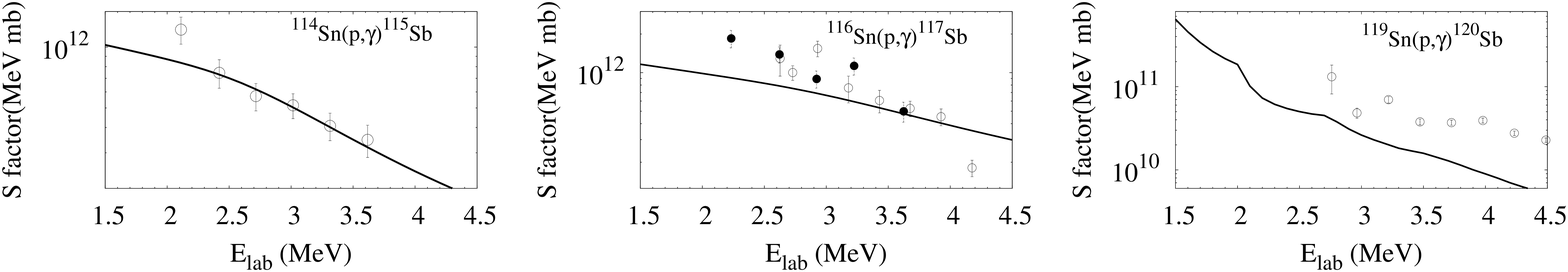}
\caption{\footnotesize Comparison of theoretical astrophysical S-factors
with experimental data for the indicated proton capture reactions. 
\label{fig:ddm3y}} 
\end{figure*}

The optical potential has been normalized for the mass region considered to fit 
the experimental observations. The DDM3Y is a purely real interaction. We have 
chosen the real part of the potential by multiplying the folding potential with a factor 0.9. The imaginary part of the potential has been chosen identical with the real one. As we will show,
it gives a reasonable agreement with the proton capture reactions in this mass 
region. Different normalizations for individual reactions may fit the observed 
data better for various reactions in this mass region. However, such a procedure
will lead to a decrease in predictive power. 
In  extending the calculation to unknown nuclei in the mass region, a single 
set of parameters will serve the purpose.

The comparison of S-factors for a number of reactions in this mass region
 are presented in Fig. \ref{fig:ddm3y}.
 We have included all experimental work where reasonable amount of data exist.
In all the figures solid lines denote the theoretical results. Open and 
solid circles with error bars denote experimental data. 
Experimental data for $^{121}$Sb and $^{123}$Sb have been taken from 
Harrisopulos {\em et al.}~\cite{hari}. In that work
experimental cross-sections were obtained from ${\gamma}$-angular 
distribution measurement with large volume HPGe detectors. 
Experimental data are from Chloupek {\em et al.}~\cite{choulpek}  
for $^{112}$Sn and $^{119}$Sn, Famiano {\em et al.}~\cite{famiano} for $^{114}$Sn and $^{116}$Sn, 
and Ozkan {\em et al.}~\cite{ozkan} for $^{116}$Sn, respectively.
Experimental data for $^{112}$Sn and $^{119}$Sn were obtained using activation technique with 
BGO detector and statistical model calculation was used to describe the results.
For $^{114}$Sn and $^{116}$Sn activation technique was used to extract the 
cross-section~\cite{famiano}. 

One can see that the present method can describe the experimental results except in a few instances. 
Of course the errors are large in some cases. Except for the proton capture 
by $^{119}$Sn, all the other reactions are reasonably described by the present 
procedure.  It is worthwhile to note that statistical model calculations by the authors failed to reproduce the cross sections for $^{119}$Sn~\cite{choulpek}.

\begin {figure*}[htp]
\centering
\includegraphics[scale=0.35]{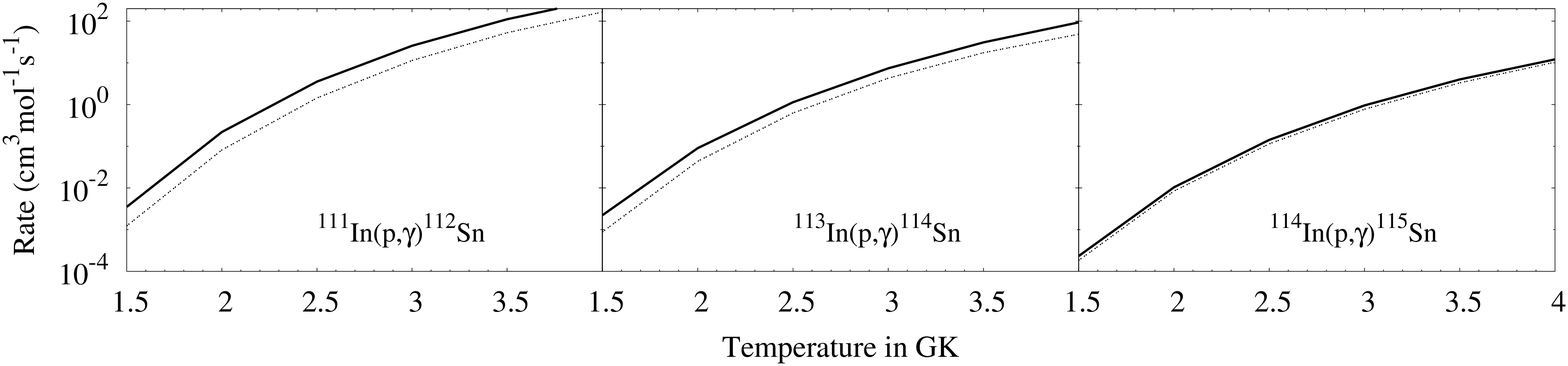}
\includegraphics[scale=0.35]{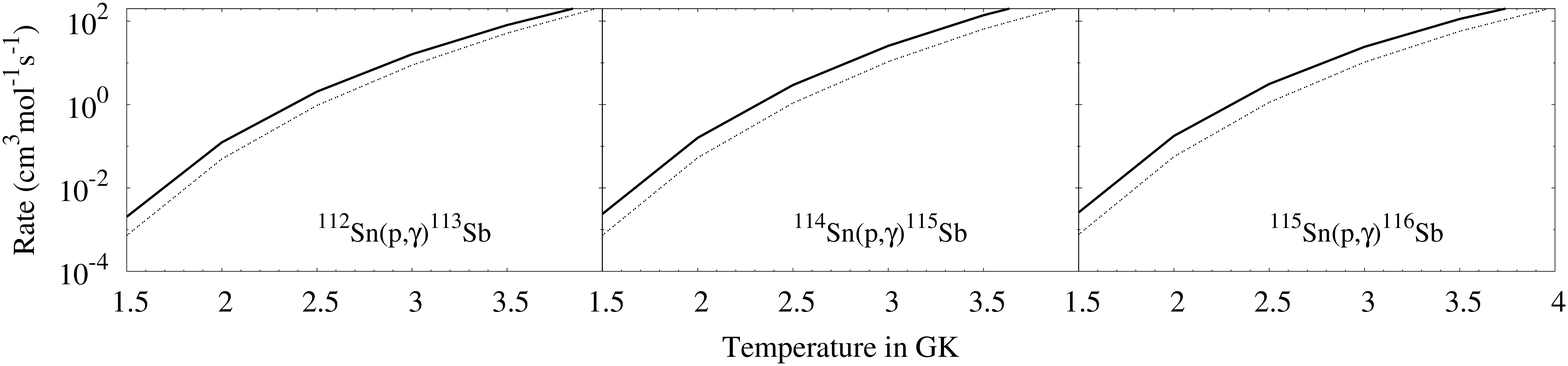}
\caption{Comparison of proton capture rates calculated from NON-SMOKER (dotted line) 
calculations and present work (solid line) for certain reactions involving p-nuclei.
\label{fig:rates}} 
\end{figure*}

We have compared the astrophysical rates calculated using NON-SMOKER 
formalism~\cite{raus1,raus2,nonsmokerweb} and our present calculation
for reactions which involve p-nuclei as target or product
to see how the present reaction rates differ
from the standard ones available in the literature. 
In Fig. \ref{fig:rates}, we plot some of the rates where the differences between
the two calculations are significant. The only exception is the  $^{114}$In$(p,\gamma)$ reaction, 
which have been shown to emphasize that in some cases, the two methods agree to a great extent.
 The temperature
$T_9$ is in GK. The data have been plotted in the temperature range 
1.5 -- 4 GK relevant to the above said Gamow window. We have used the NON-SMOKER
results for astrophysical reaction rate data with Finite Range Droplet Model
(FRDM)~\cite{nix} nuclear masses for comparison. We find that our results are
larger than the NON-SMOKER rates by a factor of 2-3 approximately in most cases.
 In some other reactions, the two 
results are very close to each other.
It should be interesting to study the effect of the present rates on the p-process.     

To summarize, a microscopic study of low energy (p,$\gamma$) reactions has been
undertaken in the mass range A=110-125. RMF calculation has been performed to 
obtain  the  nuclear density profile. 
The real and imaginary parts of the optical model potential have been obtained 
by suitably normalizing the folding potential arising out of the DDM3Y 
interaction to explain the observed reaction data. S-factor values have been
compared with experimental data. Proton capture reaction rates obtained from our 
calculation have been compared with NON-SMOKER reaction rates. 

The authors acknowledge the financial support provided by University Grants Commission, Department of Science and Technology, Alexander Von Humboldt Foundation and the University of Calcutta.

\end{document}